\documentclass[aps,prl,twocolumn,showpacs,showkeys,amsmath,superscriptaddress,floatfix]{revtex4-1}
\usepackage{graphicx}
\usepackage{epsfig}
\usepackage{dcolumn}
\usepackage{bm}
\usepackage{amsmath}
\usepackage{amsbsy}
\usepackage{amssymb}
\usepackage[usenames]{color}
\usepackage{multirow}
\usepackage{tabu}
\usepackage{soul}
\usepackage{wela}
\usepackage[T1]{fontenc}

\newcommand{\bra}[1]{\ensuremath{\langle{#1}|\,}}
\newcommand{\ket}[1]{\ensuremath{\,|{#1}\rangle}}

\begin{document}

\title{Twisted-ligth--ion interaction: the role of longitudinal fields}

\author{Guillermo\ F.\ Quinteiro}
\affiliation{ Departamento de F\'isica and IFIBA, FCEN, Universidad
de Buenos Aires, Ciudad Universitaria, Pabell\'on I, 1428 Ciudad de
Buenos Aires, Argentina}

\email{gquinteiro@df.uba.ar}

\author{Ferdinand Schmidt-Kaler}
\affiliation{Institut f\"ur Physik, Universit\"at Mainz, Staudingerweg 7, 55128 Mainz, Germany}

\author{Christian\ T.\ Schmiegelow}
\affiliation{ Departamento de F\'isica and IFIBA, FCEN, Universidad
de Buenos Aires, Ciudad Universitaria, Pabell\'on I, 1428 Ciudad de
Buenos Aires, Argentina}

\begin{abstract}

The propagation of light beams is well described using the paraxial
approximation, where  field components along the propagation
direction are usually neglected. For strongly inhomogeneous or
shaped light fields, however, this approximation may fail, leading
to intriguing variations of the light-matter interaction.  This is
the case of twisted light having opposite orbital and spin angular
momenta. We compare experimental data for the excitation of a
quadrupole transition in a single trapped $^{40}$Ca$^+$
ion by Schmiegelow et al, Nat.\ Comm.\ 7, 12998 (2016),
with a complete model where
longitudinal components of the electric field are taken into
account. Our model matches the experimental data and excludes by 11
standard deviations the approximation of complete transverse field.
This demonstrates the importance of all field components in the
interaction of twisted light with matter.
\end{abstract}

\maketitle

%
The full vector character of the electromagnetic field is
responsible for a variety of basic physical processes, such as those
occurring in near-field optics and the propagation of focused beams
close to the diffraction limit~\cite{novotny2012principles,
zhan2013vectorial}. Moreover, electromagnetic fields with components
in all possible directions play a special role in applied science.
For example, the sub-diffraction-limited focusing of radially
polarized beams producing strong longitudinal
fields~\cite{dorn2003sharper} can be used to improve material
processing~\cite{wang2008creation, meier2007material}. Also, longitudinal fields have seen application in Raman spectroscopy~\cite{saito2008z}, optical
tweezers~\cite{zhan2004trapping}, and have been used to observe
circular dichroism in non-chiral nanostructures~\cite{zambrana2014angular}.

Light carrying orbital angular momentum, known also as twisted light
(TL) or optical vortices, has introduced us to a new realm of
structured light~\cite{allen1992orb, padgett2004lig,
andrews2012angular, padgett2017orbital, franke2017light}. An unusual
property of TL has to do with the relative orientation of the
photon's angular momenta. When the orbital and spin angular momenta
are antiparallel to each other, longitudinal field components become
important.
As a result, the light-matter interaction is different for parallel
or anti-parallel momenta beams. This has been suggested in several
theoretical articles dealing with tightly-focused
TL~\cite{quinteiro2014light, quinteiro2015formulation} and
TL-related beams~\cite{bokor2005inv, iketaki2007inv,
monteiro2009ang, klimov2012mapping, zurita2002multipolar}.
Longitudinal fields in structured beams promise new applications,
such as the control of the spin state of electrons
or impurities in quantum dots~\cite{pazy2003spin,
quinteiro2014light}, and the excitation of
intersubband~\cite{sbierski2013twisted} transitions in quantum
wells~\cite{west1985first}.

For propagating fields that are not tightly focused the complexity
of the full vector model can be reduced, still retaining an
excellent description of the physics under consideration.
In the paraxial approximation~\cite{siegman1986lasers} one assumes
that the transverse profile changes slowly along the propagation
direction, here $z$. To lowest order in the ratio of wavelength to
beam waist the electric and magnetic fields have no longitudinal
component~\cite{lax1975maxwell, chen2002analyses}. Although very
common, such a strong assumption is not always correct. Theory shows
that Laguerre-Gaussian (LG) beams, the paradigmatic paraxial TL, has
a non-negligible longitudinal component when spin and orbital
angular momenta are opposite to each other~\cite{lax1975maxwell,
barnett1994orbital}.
In this article we show that the correct description of the
light-matter interaction between a single ion and a paraxial LG beam
requires the inclusion of the longitudinal electric-field
component, when the beam is in the antiparallel momenta
configuration. Thus, longitudinal field components do matter even in
the paraxial approximation, and may lead to unexpected applications,
for example to chiral quantum optics~\cite{lodahl2017chiral,
vermersch2017quantum}.

Before engaging in the light-matter model and comparison with
experiments, we briefly discuss Laguerre-Gaussian modes. These modes
are solutions of the paraxial wave equation in cylindrical
coordinates $\{r,\phi ,z\}$, and are perhaps the most studied of all
optical-vortex beams, for they can be easily produced from
conventional laser beams using computer generated holograms,
cylindrical lenses, Q-plates, etc.~\cite{andrews2011str}.
The starting point for the derivation of the electromagnetic field
is a transverse Lorentz-gauge vector potential
$\mathbf A(\mathbf r,t) = A_0 {\boldsymbol \epsilon} u(\mathbf r)
\exp(-i\omega t + i k z) + c. c.\,,$
with polarization ${\boldsymbol \epsilon}=\epsilon_x \hat{x}+
\epsilon_y \hat{y}$ and mode $u(\mathbf r)$ constructed from a
combination of a generalized Laguerre polynomial, a Gaussian
function, a polynomial and a phase factor~\cite{andrews2011str,
loudon2003theory}. Using the Lorenz condition, a scalar potential
can be deduced. From vector and scalar potentials, the electric and
magnetic fields are calculated.
In problems involving the interaction with small objects which are
centered with respect to the beam axis, the full lateral spatial
extend of the beam is irrelevant, and one can simplify the beam's
profile without significant lost of precision. This is the case for
the experiment here discussed~\cite{schmiegelow2016transfer}. The
positive part of the electric field of a circularly polarized LG
beam close to the phase singularity at $r=0$ is thus
\begin{eqnarray}
\label{Eq:E-Loudon}
    \mathbf E^{(+)}(\mathbf r)
&=&
    \frac{E_0}{\sqrt{2}}
    \left\{
        (\hat{x} + i\sigma \hat{y}) u(\mathbf r)
    \right.
\nonumber \\
&&
    \left.
        + \frac{i}{k} [\partial_x u(\mathbf r) + \partial_y u(\mathbf
r)] \hat{z}
    \right\}
    e^{i k z}
\end{eqnarray}
\begin{eqnarray}
    \text{with:   }
    u(\mathbf r)
&=&
    \frac{\sqrt{2}^{\mid\ell\mid+1}}
         {\sqrt{\pi \mid\ell\mid!}w_0^{\mid\ell\mid+1}}
    (x^2+y^2)^{\frac{\mid\ell\mid}{2}}
    e^{i \ell \arctan(y/x)}
    \nonumber
\,.
\end{eqnarray}
Here $\mathbf E(\mathbf r,t) = \mathbf E^{(+)}(\mathbf r)
e^{-i\omega t} + c. c.$,
$\sigma$ is the spin index or handness of circular polarization,
$w_0$ is the waist, $k$ is the wave number in the longitudinal
direction, and the integer $\ell$ is the orbital angular momentum
index.
In cylindrical coordinates $\{r,\varphi,z\}$ the resulting field can
be conveniently separated into transverse $\mathbf
E_\perp^{(+)}(\mathbf r)$ and longitudinal $E_z^{(+)}(\mathbf r)$
components. One obtains
\begin{subequations}
\label{Eq:E_LG}
\begin{eqnarray}
\label{Eq:Eperp}
    \mathbf E_\perp^{(+)}(\mathbf r)
&=&
    ({\hat x} + i\sigma {\hat y})
    \frac{E_ 0}{\sqrt{\pi \mid\! \ell \! \mid !}w_0}
    \left (\frac {\sqrt{2}} {w_ 0} r \right)^{\mid\ell\mid}
\nonumber \\
&& \times
    e^{i\ell\varphi} e^{i k z}
\\
\label{Eq:Ez}
    E_z^{(+)} (\mathbf r)
&=&
    -i \hat{z}
    \left(\ell\sigma-{\mid\!\ell\!\mid}\right)
    (1-\delta_{\ell,0})
    \frac{E_ 0}{\sqrt{\pi \mid\! \ell \! \mid !}w_0}
    \frac{\sqrt {2}}{w_0 k}
\nonumber \\
&& \times
    \left (\frac {\sqrt{2}} {w_ 0} r \right)^{\mid\ell\mid-1}
    e^{i(\ell+\sigma) \varphi} e^{i k z}
,
\end{eqnarray}
\end{subequations}
Here one can directly see that the longitudinal component is only
present when spin and orbital angular momenta are counter rotating,
i.\ e.\ for $\left(\ell\sigma-{\mid\!\ell\!\mid}\right)= -2
\mid\!\ell\!\mid$. In this situation is in fact crucial to take into
account the longitudinal field to accurately describe quantitatively
the atom-photon interaction. Also clear is that $E_z^{(+)} (\mathbf
r)$ is of higher order in the paraxial parameter $1/(w_0 k)$.

\begin{figure*}[t]
  \centerline{\includegraphics[scale=0.9]{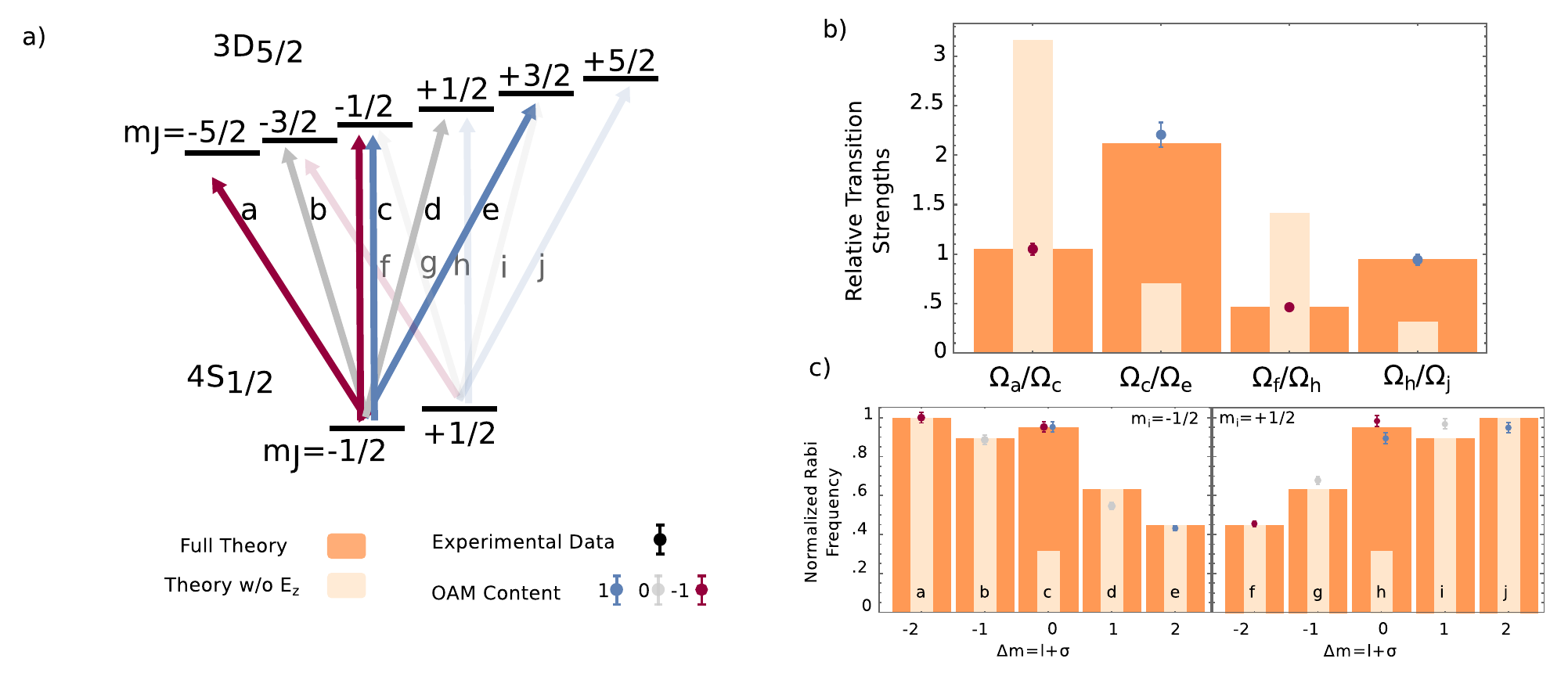}}
  \caption{
    a) Energy levels for the Zeeman split $^{40}$Ca$^+$ quadrupole
    S$_{1/2}$-D$_{5/2}$ transition. The transitions are labeled from
    $a$ to $j$ and shown in different colors indicating the photon
    orbital angular momentum content. Note that there are two different
    configurations to drive the $\Delta m =0$ transitions, either with
    $\ell$=+1 or -1.
     b) Relative transition strengths expressed as ratios of Rabi
     frequencies $\Omega$ for different transition.
      Data for the $\ell$=1 transitions (light-blue) and for the
      $\ell$=-1 (dark-purple) are plotted together with theory
      predictions. Dark orange bars show the prediction with the
      model presented in this paper where longitudinal fields are
      take into account. Light orange bars show the prediction where
      the longitudinal field is neglected.
    c) Normalization is done to a chosen transition $\Omega_a$ and
    to account for the difference in interaction strength between
    transitions driven with Gaussian or vortex beams with
    $\Omega_{norm}=\Omega_a(\delta_{|\ell|,1}+\delta_{\ell,0} kw_0)$.
    All measured transition strengths and predictions are labeled as
    in a).
  }
  \label{fig:Levels}
\end{figure*}

Intriguing features of light fields can be sensed by matter,
especially by atoms. Thus an accurate theory of light-matter
interaction which considers the full vector properties of the beam
is needed. In particular we consider the interaction of such light
fields with a single atom which is well localized with respect to
the beam size. We will consider ions or atoms cooled below they
Lamb-Dicke limit so it is well justified  to neglect linear momentum
transfer to the atom's center of mass.

The light-matter interaction for beams with $\mid\!\ell \!\mid= 0,
1$ is dominated by the electric field. In the Poincar\'e gauge, the
interaction Hamiltonian results from the scalar potential $U
(\mathbf r,t) = - \int_0^1 \mathbf r \cdot \mathbf E(u \mathbf r,t)
du$ \cite{quinteiro2017formulation}, where the explicit expressions
for the electric field Eqs.\ (\ref{Eq:Eperp})-(\ref{Eq:Ez}) are
used. Transforming to spherical coordinates $\{\rho,\varphi,\theta
\}$ and separating the interaction Hamiltonian $H_I = q U(\mathbf
r,t)$ into in-plane and $z$ sections result in the following.
For $\ell=0$:
\begin{subequations}
\label{Eq:H_0}
\begin{eqnarray}
\label{Eq:hperp_0}
    H_{I\perp}^{(+)}
&=&
    -\alpha \rho \sin(\theta)
    e^{i \sigma\varphi}
    \int_0^1  u \,  e^{i k u \rho \cos(\theta)} du
\\
\label{Eq:hz_0}
    H_{Iz}^{(+)}
&=&
    0
\,,
\end{eqnarray}
\end{subequations}
where $\alpha = \sqrt{2/\pi}q E_0/w_0^2$, with $q=-|q_e|$ the
electric charge.
For $\ell=\pm 1$:
\begin{subequations}
\label{Eq:H_1}
\begin{eqnarray}
\label{Eq:hperp_1}
    H_{I\perp}^{(+)}
&=&
    -\alpha \rho^2 \sin^2(\theta)
    e^{i (|\ell|/\ell+\sigma)\varphi}
    \int_0^1  u \,  e^{i k u \rho \cos(\theta)} du
    \hspace{+3mm}
\\
\label{Eq:hz_1}
    H_{Iz}^{(+)}
&=&
    - i \delta_{\ell,-\sigma}
    \alpha \rho \cos(\theta)
    \frac{2}{k}
    \int_0^1
        \,  e^{i k u \rho \cos(\theta)}
    du
\,.
\end{eqnarray}
\end{subequations}
The oscillating time dependence appears in $H_I = H_{I}^{(+)}e^{-i
\omega t}+ c. c.$. The positive part of $H_I$ is sufficient to
calculate transition matrix elements in the rotating wave
approximation \cite{scully1997qua}.

In the experiment a LG beam with wavelength near $\lambda=729$~nm
was focused to $w_0=2.7$~$\mu$m.  This focus size is well above the optical diffraction limit. A single ion was laser-cooled to a wavepacket position uncertainty of $\Delta r = 60$~nm, positioned at the center of the vortex field and the quadrupole $S_{1/2}$ to $D_{5/2}$ transition was excited. %

As is usual, due to the small size of the atom with respect to the
wavelength of the exciting laser, a Taylor approximation of $\exp[{i
k u \rho \cos(\theta)}]$ is justified. However, the expansion has to
be treated differently for the in-plane and $z$ sections of the
interaction. We will be concentrating on quadrupole transitions
where the initial and final state have the same parity. This causes
the transition matrix elements to vanish at lowest order in the
coordinates. The matrix elements governing these transitions are
thus those with quadratic dependence on the coordinates.

When $\ell=0$ the transitions allowed by angular momentum selection
rules are those with a change in projection of angular momentum in
the $z$ direction: $\Delta m= \pm 1$. In Fig.~\ref{fig:Levels}a
these are labelled $\{b,d\}$ and $\{g,i\}$. These transitions are
governed by Eq.\ \ref{Eq:hperp_0}. The zero order term in the expansion
of the exponential yields a Hamiltonian linear
in the coordinates, and vanishes for the quadrupole transition.
The transition is thus mediated by the following term in the expansion:
$H_{I\perp}^{(+)} \simeq - i (\alpha k \rho^2/3) \sin(\theta)
\cos(\theta) e^{i \sigma\varphi}$.

In the case $\ell=\pm 1$, in-plane and/or longitudinal interactions
may induce the transitions with no change or a change of two units
of angular momentum of the ion ($\Delta m= 0,\pm2$) depending on the
signs of the polarization ($\sigma$) and the sense of rotation of
the LG beam (the sign of $\ell$). These transitions are labelled
$\{a,c,e\}$ and $\{f,h,j\}$ in Fig.~\ref{fig:Levels}a. Following the
same reasoning as before, we must consider the first-order term
$\exp[{i k u \rho \cos(\theta)}] \simeq i k u \rho \cos(\theta)$ for
transitions induced by $H_{Iz}$, but it suffices to consider the
zero-order term $\exp[{i k u \rho \cos(\theta)}] \simeq 1$ for
$H_{I\perp}$.

It is worth noting that even though different approximations to
$\exp[{i k u \rho \cos(\theta)}]$ were used, both interaction terms
$H_{I\perp}^{(+)}$ and $H_{Iz}^{(+)}$ are of the same magnitude, a
fact that will become evident when the numerical evaluations are
performed. Moreover, the ratio of these two components does not
depend on the focus size.

We contrast out theory with recently measured interaction
strengths of different quadrupole transitions of a single trapped
Calcium ion induced by LG beams, as reported  in Ref.\
\cite{schmiegelow2016transfer}. In particular we concentrate on the
experiments where the trapped ion is illuminated by on-center
\cite{quinteiro2010ele} LG beams with different polarizations. The
LG beam is tuned to the different magnetic sublevels of the
$4S_{1/2}-3D_{5/2}$ transition which are Zeeman-split by a static
magnetic field in the $z$ direction (see Fig. \ref{fig:Levels}a).
Transitions with different change in atom angular momentum
$\Delta m = -2,-1,0,1,2$ were excited with all combinations of
spin and orbital angular momentum of the beam $m_p = \sigma + \ell$.
Here, we restrict to the cases where the total angular momentum
of the photon matches the change of angular momentum of the atom
$\Delta m = m_p$. The experiments
determined the strength of each transition by measuring the
frequency of its Rabi oscillations by state-dependent fluorescence \cite{sauter1986observation}. The Rabi frequencies are proportional to the matrix element of the transition.


To complete the evaluation of matrix elements and comparison with
experimental data we follow this algebraic
manipulations~\cite{schmiegelow2012light}: we
$i)$ rewrite the interactions in terms of products of spherical
tensors using $ \sin(\theta)\exp(\pm i \varphi)=\mp \sqrt{8\pi/3}
\,T_{\pm 1}^{(1)}$ and $\cos(\theta)=\sqrt{{4\pi}/{3}} \,
T_{0}^{(1)}$;
$ii)$ convert products of lower-rank tensors $\{T_{q_1}^{(k_1)},
T_{q_2}^{(k_2)}\}$ into sums of single higher-rank tensors
$T_{q}^{(k)}$ times Clebsch-Gordan coefficients with
$T_{q_1}^{(k_1)} T_{q_2}^{(k_2)} = \sum_{kq} T_{q}^{(k)} (kq|k_1 k_2
q_1 q_2)$;
$iii)$ regroup interaction terms by orbital and spin angular
momentum $H_{I} (\ell,\sigma)$;
$iv)$ use the Wigner-Eckart theorem
to calculate matrix elements $\bra{3D_{5/2},
m_{j'}}H_{I}(\ell,\sigma)\ket{4S_{1/2}, m_j}$ (with and without
$H_{Iz}$) and their ratios.

To exemplify the previous procedure we outline the particular
case of transitions $c$ and $e$.  These are driven by a field with
$\ell=1$, with $\sigma=-1$ and $\sigma=1$, with corresponding matrix
elements $M_{c}=\bra{3D_{5/2}, -1/2}H_{I}(1,-1)\ket{4S_{1/2}, -1/2}$ and $M_{e}=\bra{3D_{5/2}, 3/2}H_{I}(1,1)\ket{4S_{1/2}, -1/2}$.
Steps $i)$-$iii)$ above lead to
\begin{subequations}
\label{Eq:H2}
\begin{eqnarray}
\label{Eq:s1}
    H_{I}(1,1)
&=&
    -\frac{4\pi}{3}\alpha \, \rho^2 \, T_{+2}^{(2)}
\\
\label{Eq:s-1}
    H_{I}(1,-1)
&=&
    \frac{4\pi}{3}\alpha \, \rho^2 \,
    \left[
    \frac{n\, T_{0}^{(2)}}{\sqrt{6}}
    - \frac{T_{0}^{(1)}}{\sqrt{2}}
    \right]
\,,
\end{eqnarray}
\end{subequations}
where we introduced the integer $n$ to encopass both cases in one formula: $n=3$ for the full Hamiltonian
$H_{I}^{(+)}=H_{I\perp}^{(+)}+H_{Iz}^{(+)}$ and $n=1$ for the case
where we omit the logitudinal field $H_{I}^{(+)}=H_{I\perp}^{(+)}$.
The Wigner-Eckart theorem requires the calculation of Clebsch-Gordan
coefficients, coupling tensor indices $(k,q)$, initial-state indices
$(J,m_j)$ and final-state indices $(J',m_{j'})$. In our case, these
Clebsch-Gordan coefficients are nonzero only for tensors of rank
$k=2$.
Moreover, given that all matrix elements share the same initial and
final radial and total angular momentum quantum numbers, the reduced
matrix element is the same for all $M_\gamma$. Then, the ratio of
matrix elements depend on both, the specific prefactors arising from
the decomposition of tensor products and the Clebsch-Gordan
coefficients arising from the Wigner-Eckart theorem. The predictions
for the ratios without and with $E_z$ are
\begin{eqnarray}
\label{Eq:ratios}
    \text{case}~~ E_z=0: {\text ~~~~}
    \bigg \lvert
    \frac{M_{c}}{M_{e}}
    \bigg \rvert
&=&
    \frac{1}{\sqrt{2}} \simeq 0.707
\\
    \text{case}~~ E_z\neq 0: {\text ~~~~}
    \bigg \lvert
    \frac{M_{c}}{M_{e}}
    \bigg \rvert
&=&
    \frac{3}{\sqrt{2}} \simeq 2.123
\,.
\end{eqnarray}
These are to be compared to the ratio of power-normalized Rabi
frequencies $\Omega_\gamma\propto M_\gamma $. The experimental value
for the ratio $\Omega_c/\Omega_e=2.21\pm 0.13$, see Fig.\ \ref{fig:Levels}b. This matches the theory
that includes the excitation by the longitudinal field component, but is off from the theory with $E_z=0$ by more than 11 standard deviations.

In this spirit, we calculate all other matrix elements and
compare them to the experimental data for transitions induced by opposite polarizations with the same LG beam, either with $\ell=1$ (light-blue) or $\ell=-1$ (dark-purple), and for the two possible initial states, see Fig.\
\ref{fig:Levels}b. Together with the data, we plot the predictions with and without the longitudinal field. We see that the results match the theory with the full vector field. The incomplete theory is off in all cases by more than 11 standard deviations.


%
%
%

The interaction of matter with strong inhomogeneous fields shows
complex features and exposes new physics. Our results on the
interaction of Laguerre-Gaussian modes with single ions demonstrate
that an extremely good quantitative match between predictions and
measurements is achieved only if an improved version of the paraxial
approximation is used, where the longitudinal field is taken into
account.
In contrast, a theory that lacks the longitudinal component misses
the experimental values for up to 11 standard deviations.

The relative contributions to the interaction strength of
the longitudinal and transverse components in a beam with
$\ell=1$ does not depend on the beam waist $w_0$, it is only
determined by the relative sense of spin and orbital angular
momenta.

Longitudinal fields are important in quantum optics setups, such as
trapped ions, but also well beyond as for instance for neutral
atoms, molecules and quantum dots. Our finding of the significance
of longitudinal fields may pave new ways to manipulate matter for
spintronics, nanophotonics and quantum information.
But besides possible new uses, countless applications thought to be
implemented by more technically-demanding tools of near-field optics
or tight focusing may become addressable by simple beams of twisted
light. One important example is the chiral light-matter interaction
from the longitudinal component of light propagating in
sub-wavelength waveguides to produce directional coupling in
nano-photonic chips~\cite{lodahl2017chiral}.
In the future, we plan exploring chiral coupling effects from
free-propagating LG light beams, specifically for the different
platforms of trapped ions or colloidal quantum dots.


%


\begin{thebibliography}{37}%
\makeatletter
\providecommand \@ifxundefined [1]{%
 \@ifx{#1\undefined}
}%
\providecommand \@ifnum [1]{%
 \ifnum #1\expandafter \@firstoftwo
 \else \expandafter \@secondoftwo
 \fi
}%
\providecommand \@ifx [1]{%
 \ifx #1\expandafter \@firstoftwo
 \else \expandafter \@secondoftwo
 \fi
}%
\providecommand \natexlab [1]{#1}%
\providecommand \enquote  [1]{``#1''}%
\providecommand \bibnamefont  [1]{#1}%
\providecommand \bibfnamefont [1]{#1}%
\providecommand \citenamefont [1]{#1}%
\providecommand \href@noop [0]{\@secondoftwo}%
\providecommand \href [0]{\begingroup \@sanitize@url \@href}%
\providecommand \@href[1]{\@@startlink{#1}\@@href}%
\providecommand \@@href[1]{\endgroup#1\@@endlink}%
\providecommand \@sanitize@url [0]{\catcode `\\12\catcode
`\$12\catcode
  `\&12\catcode `\#12\catcode `\^12\catcode `\_12\catcode `\%12\relax}%
\providecommand \@@startlink[1]{}%
\providecommand \@@endlink[0]{}%
\providecommand \url  [0]{\begingroup\@sanitize@url \@url }%
\providecommand \@url [1]{\endgroup\@href {#1}{\urlprefix }}%
\providecommand \urlprefix  [0]{URL }%
\providecommand \Eprint [0]{\href }%
\providecommand \doibase [0]{http://dx.doi.org/}%
\providecommand \selectlanguage [0]{\@gobble}%
\providecommand \bibinfo  [0]{\@secondoftwo}%
\providecommand \bibfield  [0]{\@secondoftwo}%
\providecommand \translation [1]{[#1]}%
\providecommand \BibitemOpen [0]{}%
\providecommand \bibitemStop [0]{}%
\providecommand \bibitemNoStop [0]{.\EOS\space}%
\providecommand \EOS [0]{\spacefactor3000\relax}%
\providecommand \BibitemShut  [1]{\csname bibitem#1\endcsname}%
\let\auto@bib@innerbib\@empty
\bibitem [{\citenamefont {Novotny}\ and\ \citenamefont
  {Hecht}(2012)}]{novotny2012principles}%
  \BibitemOpen
  \bibfield  {author} {\bibinfo {author} {\bibfnamefont {L.}~\bibnamefont
  {Novotny}}\ and\ \bibinfo {author} {\bibfnamefont {B.}~\bibnamefont
  {Hecht}},\ }\href@noop {} {\emph {\bibinfo {title} {Principles of
  nano-optics}}}\ (\bibinfo  {publisher} {Cambridge university press},\
  \bibinfo {year} {2012})\BibitemShut {NoStop}%
\bibitem [{\citenamefont {Zhan}(2013)}]{zhan2013vectorial}%
  \BibitemOpen
  \bibfield  {author} {\bibinfo {author} {\bibfnamefont {Q.}~\bibnamefont
  {Zhan}},\ }\href@noop {} {\emph {\bibinfo {title} {Vectorial optical fields:
  Fundamentals and applications}}}\ (\bibinfo  {publisher} {World Scientific},\
  \bibinfo {year} {2013})\BibitemShut {NoStop}%
\bibitem [{\citenamefont {Dorn}\ \emph {et~al.}(2003)\citenamefont {Dorn},
  \citenamefont {Quabis},\ and\ \citenamefont {Leuchs}}]{dorn2003sharper}%
  \BibitemOpen
  \bibfield  {author} {\bibinfo {author} {\bibfnamefont {R.}~\bibnamefont
  {Dorn}}, \bibinfo {author} {\bibfnamefont {S.}~\bibnamefont {Quabis}}, \ and\
  \bibinfo {author} {\bibfnamefont {G.}~\bibnamefont {Leuchs}},\ }\href@noop {}
  {\bibfield  {journal} {\bibinfo  {journal} {Phys.\ Rev.\ Lett.}\ }\textbf
  {\bibinfo {volume} {91}},\ \bibinfo {pages} {233901} (\bibinfo {year}
  {2003})}\BibitemShut {NoStop}%
\bibitem [{\citenamefont {Wang}\ \emph {et~al.}(2008)\citenamefont {Wang},
  \citenamefont {Shi}, \citenamefont {Lukyanchuk}, \citenamefont {Sheppard},\
  and\ \citenamefont {Chong}}]{wang2008creation}%
  \BibitemOpen
  \bibfield  {author} {\bibinfo {author} {\bibfnamefont {H.}~\bibnamefont
  {Wang}}, \bibinfo {author} {\bibfnamefont {L.}~\bibnamefont {Shi}}, \bibinfo
  {author} {\bibfnamefont {B.}~\bibnamefont {Lukyanchuk}}, \bibinfo {author}
  {\bibfnamefont {C.}~\bibnamefont {Sheppard}}, \ and\ \bibinfo {author}
  {\bibfnamefont {C.~T.}\ \bibnamefont {Chong}},\ }\href@noop {} {\bibfield
  {journal} {\bibinfo  {journal} {Nature Photonics}\ }\textbf {\bibinfo
  {volume} {2}},\ \bibinfo {pages} {501} (\bibinfo {year} {2008})}\BibitemShut
  {NoStop}%
\bibitem [{\citenamefont {Meier}\ \emph {et~al.}(2007)\citenamefont {Meier},
  \citenamefont {Romano},\ and\ \citenamefont {Feurer}}]{meier2007material}%
  \BibitemOpen
  \bibfield  {author} {\bibinfo {author} {\bibfnamefont {M.}~\bibnamefont
  {Meier}}, \bibinfo {author} {\bibfnamefont {V.}~\bibnamefont {Romano}}, \
  and\ \bibinfo {author} {\bibfnamefont {T.}~\bibnamefont {Feurer}},\
  }\href@noop {} {\bibfield  {journal} {\bibinfo  {journal} {Applied Physics A:
  Materials Science \& Processing}\ }\textbf {\bibinfo {volume} {86}},\
  \bibinfo {pages} {329} (\bibinfo {year} {2007})}\BibitemShut {NoStop}%
\bibitem [{\citenamefont {Saito}\ \emph {et~al.}(2008)\citenamefont {Saito},
  \citenamefont {Kobayashi}, \citenamefont {Hiraga}, \citenamefont {Fujita},
  \citenamefont {Kawano}, \citenamefont {Smith}, \citenamefont {Inouye},\ and\
  \citenamefont {Kawata}}]{saito2008z}%
  \BibitemOpen
  \bibfield  {author} {\bibinfo {author} {\bibfnamefont {Y.}~\bibnamefont
  {Saito}}, \bibinfo {author} {\bibfnamefont {M.}~\bibnamefont {Kobayashi}},
  \bibinfo {author} {\bibfnamefont {D.}~\bibnamefont {Hiraga}}, \bibinfo
  {author} {\bibfnamefont {K.}~\bibnamefont {Fujita}}, \bibinfo {author}
  {\bibfnamefont {S.}~\bibnamefont {Kawano}}, \bibinfo {author} {\bibfnamefont
  {N.}~\bibnamefont {Smith}}, \bibinfo {author} {\bibfnamefont
  {Y.}~\bibnamefont {Inouye}}, \ and\ \bibinfo {author} {\bibfnamefont
  {S.}~\bibnamefont {Kawata}},\ }\href@noop {} {\bibfield  {journal} {\bibinfo
  {journal} {Journal of Raman Spectroscopy}\ }\textbf {\bibinfo {volume}
  {39}},\ \bibinfo {pages} {1643} (\bibinfo {year} {2008})}\BibitemShut
  {NoStop}%
\bibitem [{\citenamefont {Zhan}(2004)}]{zhan2004trapping}%
  \BibitemOpen
  \bibfield  {author} {\bibinfo {author} {\bibfnamefont {Q.}~\bibnamefont
  {Zhan}},\ }\href@noop {} {\bibfield  {journal} {\bibinfo  {journal} {Optics
  express}\ }\textbf {\bibinfo {volume} {12}},\ \bibinfo {pages} {3377}
  (\bibinfo {year} {2004})}\BibitemShut {NoStop}%
\bibitem [{\citenamefont {Zambrana-Puyalto}\ \emph {et~al.}(2014)\citenamefont
  {Zambrana-Puyalto}, \citenamefont {Vidal},\ and\ \citenamefont
  {Molina-Terriza}}]{zambrana2014angular}%
  \BibitemOpen
  \bibfield  {author} {\bibinfo {author} {\bibfnamefont {X.}~\bibnamefont
  {Zambrana-Puyalto}}, \bibinfo {author} {\bibfnamefont {X.}~\bibnamefont
  {Vidal}}, \ and\ \bibinfo {author} {\bibfnamefont {G.}~\bibnamefont
  {Molina-Terriza}},\ }\href@noop {} {\bibfield  {journal} {\bibinfo  {journal}
  {Nature communications}\ }\textbf {\bibinfo {volume} {5}} (\bibinfo {year}
  {2014})}\BibitemShut {NoStop}%
\bibitem [{\citenamefont {Allen}\ \emph {et~al.}(1992)\citenamefont {Allen},
  \citenamefont {Beijersbergen}, \citenamefont {Spreeuw},\ and\ \citenamefont
  {Woerdman}}]{allen1992orb}%
  \BibitemOpen
  \bibfield  {author} {\bibinfo {author} {\bibfnamefont {L.}~\bibnamefont
  {Allen}}, \bibinfo {author} {\bibfnamefont {M.~W.}\ \bibnamefont
  {Beijersbergen}}, \bibinfo {author} {\bibfnamefont {R.~J.~C.}\ \bibnamefont
  {Spreeuw}}, \ and\ \bibinfo {author} {\bibfnamefont {J.~P.}\ \bibnamefont
  {Woerdman}},\ }\href@noop {} {\bibfield  {journal} {\bibinfo  {journal}
  {Phys.\ Rev.\ A}\ }\textbf {\bibinfo {volume} {45}},\ \bibinfo {pages} {8185}
  (\bibinfo {year} {1992})}\BibitemShut {NoStop}%
\bibitem [{\citenamefont {Padgett}\ \emph {et~al.}(2004)\citenamefont
  {Padgett}, \citenamefont {Courtial},\ and\ \citenamefont
  {Allen}}]{padgett2004lig}%
  \BibitemOpen
  \bibfield  {author} {\bibinfo {author} {\bibfnamefont {M.}~\bibnamefont
  {Padgett}}, \bibinfo {author} {\bibfnamefont {J.}~\bibnamefont {Courtial}}, \
  and\ \bibinfo {author} {\bibfnamefont {L.}~\bibnamefont {Allen}},\
  }\href@noop {} {\bibfield  {journal} {\bibinfo  {journal} {Physics Today}\
  }\textbf {\bibinfo {volume} {57}},\ \bibinfo {pages} {35} (\bibinfo {year}
  {2004})}\BibitemShut {NoStop}%
\bibitem [{\citenamefont {Andrews}\ and\ \citenamefont
  {Babiker}(2012)}]{andrews2012angular}%
  \BibitemOpen
  \bibfield  {author} {\bibinfo {author} {\bibfnamefont {D.~L.}\ \bibnamefont
  {Andrews}}\ and\ \bibinfo {author} {\bibfnamefont {M.}~\bibnamefont
  {Babiker}},\ }\href@noop {} {\emph {\bibinfo {title} {The angular momentum of
  light}}}\ (\bibinfo  {publisher} {Cambridge University Press},\ \bibinfo
  {year} {2012})\BibitemShut {NoStop}%
\bibitem [{\citenamefont {Padgett}(2017)}]{padgett2017orbital}%
  \BibitemOpen
  \bibfield  {author} {\bibinfo {author} {\bibfnamefont {M.~J.}\ \bibnamefont
  {Padgett}},\ }\href@noop {} {\bibfield  {journal} {\bibinfo  {journal}
  {Optics Express}\ }\textbf {\bibinfo {volume} {25}},\ \bibinfo {pages}
  {11265} (\bibinfo {year} {2017})}\BibitemShut {NoStop}%
\bibitem [{\citenamefont {Franke-Arnold}\ and\ \citenamefont
  {Radwell}(2017)}]{franke2017light}%
  \BibitemOpen
  \bibfield  {author} {\bibinfo {author} {\bibfnamefont {S.}~\bibnamefont
  {Franke-Arnold}}\ and\ \bibinfo {author} {\bibfnamefont {N.}~\bibnamefont
  {Radwell}},\ }\href@noop {} {\bibfield  {journal} {\bibinfo  {journal}
  {Optics and Photonics News}\ }\textbf {\bibinfo {volume} {28}},\ \bibinfo
  {pages} {28} (\bibinfo {year} {2017})}\BibitemShut {NoStop}%
\bibitem [{\citenamefont {Quinteiro}\ and\ \citenamefont
  {Kuhn}(2014)}]{quinteiro2014light}%
  \BibitemOpen
  \bibfield  {author} {\bibinfo {author} {\bibfnamefont {G.~F.}\ \bibnamefont
  {Quinteiro}}\ and\ \bibinfo {author} {\bibfnamefont {T.}~\bibnamefont
  {Kuhn}},\ }\href@noop {} {\bibfield  {journal} {\bibinfo  {journal} {Phys.
  Rev. B}\ }\textbf {\bibinfo {volume} {90}},\ \bibinfo {pages} {115401}
  (\bibinfo {year} {2014})}\BibitemShut {NoStop}%
\bibitem [{\citenamefont {Quinteiro}\ \emph {et~al.}(2015)\citenamefont
  {Quinteiro}, \citenamefont {Reiter},\ and\ \citenamefont
  {Kuhn}}]{quinteiro2015formulation}%
  \BibitemOpen
  \bibfield  {author} {\bibinfo {author} {\bibfnamefont {G.}~\bibnamefont
  {Quinteiro}}, \bibinfo {author} {\bibfnamefont {D.}~\bibnamefont {Reiter}}, \
  and\ \bibinfo {author} {\bibfnamefont {T.}~\bibnamefont {Kuhn}},\ }\href@noop
  {} {\bibfield  {journal} {\bibinfo  {journal} {Physical Review A}\ }\textbf
  {\bibinfo {volume} {91}},\ \bibinfo {pages} {033808} (\bibinfo {year}
  {2015})}\BibitemShut {NoStop}%
\bibitem [{\citenamefont {Bokor}\ \emph {et~al.}(2005)\citenamefont {Bokor},
  \citenamefont {Iketaki}, \citenamefont {Watanabe},\ and\ \citenamefont
  {Fujii}}]{bokor2005inv}%
  \BibitemOpen
  \bibfield  {author} {\bibinfo {author} {\bibfnamefont {N.}~\bibnamefont
  {Bokor}}, \bibinfo {author} {\bibfnamefont {Y.}~\bibnamefont {Iketaki}},
  \bibinfo {author} {\bibfnamefont {T.}~\bibnamefont {Watanabe}}, \ and\
  \bibinfo {author} {\bibfnamefont {M.}~\bibnamefont {Fujii}},\ }\href@noop {}
  {\bibfield  {journal} {\bibinfo  {journal} {Optics Express}\ }\textbf
  {\bibinfo {volume} {13}},\ \bibinfo {pages} {10440} (\bibinfo {year}
  {2005})}\BibitemShut {NoStop}%
\bibitem [{\citenamefont {Iketaki}\ \emph {et~al.}(2007)\citenamefont
  {Iketaki}, \citenamefont {Watanabe}, \citenamefont {Bokor},\ and\
  \citenamefont {Fujii}}]{iketaki2007inv}%
  \BibitemOpen
  \bibfield  {author} {\bibinfo {author} {\bibfnamefont {Y.}~\bibnamefont
  {Iketaki}}, \bibinfo {author} {\bibfnamefont {T.}~\bibnamefont {Watanabe}},
  \bibinfo {author} {\bibfnamefont {N.}~\bibnamefont {Bokor}}, \ and\ \bibinfo
  {author} {\bibfnamefont {M.}~\bibnamefont {Fujii}},\ }\href@noop {}
  {\bibfield  {journal} {\bibinfo  {journal} {Opt.\ Lett.}\ }\textbf {\bibinfo
  {volume} {32}},\ \bibinfo {pages} {2357} (\bibinfo {year}
  {2007})}\BibitemShut {NoStop}%
\bibitem [{\citenamefont {Monteiro}\ \emph {et~al.}(2009)\citenamefont
  {Monteiro}, \citenamefont {Neto},\ and\ \citenamefont
  {Nussenzveig}}]{monteiro2009ang}%
  \BibitemOpen
  \bibfield  {author} {\bibinfo {author} {\bibfnamefont {P.~B.}\ \bibnamefont
  {Monteiro}}, \bibinfo {author} {\bibfnamefont {P.~A.~M.}\ \bibnamefont
  {Neto}}, \ and\ \bibinfo {author} {\bibfnamefont {H.~M.}\ \bibnamefont
  {Nussenzveig}},\ }\href@noop {} {\bibfield  {journal} {\bibinfo  {journal}
  {Phys.\ Rev.\ A}\ }\textbf {\bibinfo {volume} {79}},\ \bibinfo {pages}
  {033830} (\bibinfo {year} {2009})}\BibitemShut {NoStop}%
\bibitem [{\citenamefont {Klimov}\ \emph {et~al.}(2012)\citenamefont {Klimov},
  \citenamefont {Bloch}, \citenamefont {Ducloy},\ and\ \citenamefont
  {Leite}}]{klimov2012mapping}%
  \BibitemOpen
  \bibfield  {author} {\bibinfo {author} {\bibfnamefont {V.~V.}\ \bibnamefont
  {Klimov}}, \bibinfo {author} {\bibfnamefont {D.}~\bibnamefont {Bloch}},
  \bibinfo {author} {\bibfnamefont {M.}~\bibnamefont {Ducloy}}, \ and\ \bibinfo
  {author} {\bibfnamefont {J.~R.}\ \bibnamefont {Leite}},\ }\href@noop {}
  {\bibfield  {journal} {\bibinfo  {journal} {Phys.\ Rev.\ A}\ }\textbf
  {\bibinfo {volume} {85}},\ \bibinfo {pages} {053834} (\bibinfo {year}
  {2012})}\BibitemShut {NoStop}%
\bibitem [{\citenamefont {Zurita-S{\'a}nchez}\ and\ \citenamefont
  {Novotny}(2002)}]{zurita2002multipolar}%
  \BibitemOpen
  \bibfield  {author} {\bibinfo {author} {\bibfnamefont {J.~R.}\ \bibnamefont
  {Zurita-S{\'a}nchez}}\ and\ \bibinfo {author} {\bibfnamefont
  {L.}~\bibnamefont {Novotny}},\ }\href@noop {} {\bibfield  {journal} {\bibinfo
   {journal} {JOSA B}\ }\textbf {\bibinfo {volume} {19}},\ \bibinfo {pages}
  {2722} (\bibinfo {year} {2002})}\BibitemShut {NoStop}%
\bibitem [{\citenamefont {Pazy}\ \emph {et~al.}(2003)\citenamefont {Pazy},
  \citenamefont {Biolatti}, \citenamefont {Calarco}, \citenamefont {D'amico},
  \citenamefont {Zanardi}, \citenamefont {Rossi},\ and\ \citenamefont
  {Zoller}}]{pazy2003spin}%
  \BibitemOpen
  \bibfield  {author} {\bibinfo {author} {\bibfnamefont {E.}~\bibnamefont
  {Pazy}}, \bibinfo {author} {\bibfnamefont {E.}~\bibnamefont {Biolatti}},
  \bibinfo {author} {\bibfnamefont {T.}~\bibnamefont {Calarco}}, \bibinfo
  {author} {\bibfnamefont {I.}~\bibnamefont {D'amico}}, \bibinfo {author}
  {\bibfnamefont {P.}~\bibnamefont {Zanardi}}, \bibinfo {author} {\bibfnamefont
  {F.}~\bibnamefont {Rossi}}, \ and\ \bibinfo {author} {\bibfnamefont
  {P.}~\bibnamefont {Zoller}},\ }\href@noop {} {\bibfield  {journal} {\bibinfo
  {journal} {EPL (Europhysics Letters)}\ }\textbf {\bibinfo {volume} {62}},\
  \bibinfo {pages} {175} (\bibinfo {year} {2003})}\BibitemShut {NoStop}%
\bibitem [{\citenamefont {Sbierski}\ \emph {et~al.}(2013)\citenamefont
  {Sbierski}, \citenamefont {Quinteiro},\ and\ \citenamefont
  {Tamborenea}}]{sbierski2013twisted}%
  \BibitemOpen
  \bibfield  {author} {\bibinfo {author} {\bibfnamefont {B.}~\bibnamefont
  {Sbierski}}, \bibinfo {author} {\bibfnamefont {G.}~\bibnamefont {Quinteiro}},
  \ and\ \bibinfo {author} {\bibfnamefont {P.}~\bibnamefont {Tamborenea}},\
  }\href@noop {} {\bibfield  {journal} {\bibinfo  {journal} {Journal of
  Physics: Condensed Matter}\ }\textbf {\bibinfo {volume} {25}},\ \bibinfo
  {pages} {385301} (\bibinfo {year} {2013})}\BibitemShut {NoStop}%
\bibitem [{\citenamefont {West}\ and\ \citenamefont
  {Eglash}(1985)}]{west1985first}%
  \BibitemOpen
  \bibfield  {author} {\bibinfo {author} {\bibfnamefont {L.}~\bibnamefont
  {West}}\ and\ \bibinfo {author} {\bibfnamefont {S.}~\bibnamefont {Eglash}},\
  }\href@noop {} {\bibfield  {journal} {\bibinfo  {journal} {Applied Physics
  Letters}\ }\textbf {\bibinfo {volume} {46}},\ \bibinfo {pages} {1156}
  (\bibinfo {year} {1985})}\BibitemShut {NoStop}%
\bibitem [{\citenamefont {Siegman}(1986)}]{siegman1986lasers}%
  \BibitemOpen
  \bibfield  {author} {\bibinfo {author} {\bibfnamefont {A.~E.}\ \bibnamefont
  {Siegman}},\ }\href@noop {} {\emph {\bibinfo {title} {Lasers}}}\ (\bibinfo
  {publisher} {University Science Books, Mill Valley, CA},\ \bibinfo {year}
  {1986})\ p.\ \bibinfo {pages} {276}\BibitemShut {NoStop}%
\bibitem [{\citenamefont {Lax}\ \emph {et~al.}(1975)\citenamefont {Lax},
  \citenamefont {Louisell},\ and\ \citenamefont {McKnight}}]{lax1975maxwell}%
  \BibitemOpen
  \bibfield  {author} {\bibinfo {author} {\bibfnamefont {M.}~\bibnamefont
  {Lax}}, \bibinfo {author} {\bibfnamefont {W.~H.}\ \bibnamefont {Louisell}}, \
  and\ \bibinfo {author} {\bibfnamefont {W.~B.}\ \bibnamefont {McKnight}},\
  }\href@noop {} {\bibfield  {journal} {\bibinfo  {journal} {Physical Review
  A}\ }\textbf {\bibinfo {volume} {11}},\ \bibinfo {pages} {1365} (\bibinfo
  {year} {1975})}\BibitemShut {NoStop}%
\bibitem [{\citenamefont {Chen}\ \emph {et~al.}(2002)\citenamefont {Chen},
  \citenamefont {Konkola}, \citenamefont {Ferrera}, \citenamefont {Heilmann},\
  and\ \citenamefont {Schattenburg}}]{chen2002analyses}%
  \BibitemOpen
  \bibfield  {author} {\bibinfo {author} {\bibfnamefont {C.~G.}\ \bibnamefont
  {Chen}}, \bibinfo {author} {\bibfnamefont {P.~T.}\ \bibnamefont {Konkola}},
  \bibinfo {author} {\bibfnamefont {J.}~\bibnamefont {Ferrera}}, \bibinfo
  {author} {\bibfnamefont {R.~K.}\ \bibnamefont {Heilmann}}, \ and\ \bibinfo
  {author} {\bibfnamefont {M.~L.}\ \bibnamefont {Schattenburg}},\ }\href@noop
  {} {\bibfield  {journal} {\bibinfo  {journal} {JOSA A}\ }\textbf {\bibinfo
  {volume} {19}},\ \bibinfo {pages} {404} (\bibinfo {year} {2002})}\BibitemShut
  {NoStop}%
\bibitem [{\citenamefont {Barnett}\ and\ \citenamefont
  {Allen}(1994)}]{barnett1994orbital}%
  \BibitemOpen
  \bibfield  {author} {\bibinfo {author} {\bibfnamefont {S.~M.}\ \bibnamefont
  {Barnett}}\ and\ \bibinfo {author} {\bibfnamefont {L.}~\bibnamefont
  {Allen}},\ }\href@noop {} {\bibfield  {journal} {\bibinfo  {journal} {Optics
  communications}\ }\textbf {\bibinfo {volume} {110}},\ \bibinfo {pages} {670}
  (\bibinfo {year} {1994})}\BibitemShut {NoStop}%
\bibitem [{\citenamefont {Lodahl}\ \emph {et~al.}(2017)\citenamefont {Lodahl},
  \citenamefont {Mahmoodian}, \citenamefont {Stobbe}, \citenamefont
  {Rauschenbeutel}, \citenamefont {Schneeweiss}, \citenamefont {Volz},
  \citenamefont {Pichler},\ and\ \citenamefont {Zoller}}]{lodahl2017chiral}%
  \BibitemOpen
  \bibfield  {author} {\bibinfo {author} {\bibfnamefont {P.}~\bibnamefont
  {Lodahl}}, \bibinfo {author} {\bibfnamefont {S.}~\bibnamefont {Mahmoodian}},
  \bibinfo {author} {\bibfnamefont {S.}~\bibnamefont {Stobbe}}, \bibinfo
  {author} {\bibfnamefont {A.}~\bibnamefont {Rauschenbeutel}}, \bibinfo
  {author} {\bibfnamefont {P.}~\bibnamefont {Schneeweiss}}, \bibinfo {author}
  {\bibfnamefont {J.}~\bibnamefont {Volz}}, \bibinfo {author} {\bibfnamefont
  {H.}~\bibnamefont {Pichler}}, \ and\ \bibinfo {author} {\bibfnamefont
  {P.}~\bibnamefont {Zoller}},\ }\href@noop {} {\bibfield  {journal} {\bibinfo
  {journal} {Nature}\ }\textbf {\bibinfo {volume} {541}},\ \bibinfo {pages}
  {473} (\bibinfo {year} {2017})}\BibitemShut {NoStop}%
\bibitem [{\citenamefont {Vermersch}\ \emph {et~al.}(2017)\citenamefont
  {Vermersch}, \citenamefont {Guimond}, \citenamefont {Pichler},\ and\
  \citenamefont {Zoller}}]{vermersch2017quantum}%
  \BibitemOpen
  \bibfield  {author} {\bibinfo {author} {\bibfnamefont {B.}~\bibnamefont
  {Vermersch}}, \bibinfo {author} {\bibfnamefont {P.-O.}\ \bibnamefont
  {Guimond}}, \bibinfo {author} {\bibfnamefont {H.}~\bibnamefont {Pichler}}, \
  and\ \bibinfo {author} {\bibfnamefont {P.}~\bibnamefont {Zoller}},\
  }\href@noop {} {\bibfield  {journal} {\bibinfo  {journal} {Physical Review
  Letters}\ }\textbf {\bibinfo {volume} {118}},\ \bibinfo {pages} {133601}
  (\bibinfo {year} {2017})}\BibitemShut {NoStop}%
\bibitem [{\citenamefont {Andrews}(2008)}]{andrews2011str}%
  \BibitemOpen
  \bibfield  {author} {\bibinfo {author} {\bibfnamefont {D.~L.}\ \bibnamefont
  {Andrews}},\ }\href@noop {} {\emph {\bibinfo {title} {Structured light and
  its applications: An introduction to phase-structured beams and nanoscale
  optical forces}}}\ (\bibinfo  {publisher} {Academic Press},\ \bibinfo {year}
  {2008})\BibitemShut {NoStop}%
\bibitem [{\citenamefont {Loudon}(2003)}]{loudon2003theory}%
  \BibitemOpen
  \bibfield  {author} {\bibinfo {author} {\bibfnamefont {R.}~\bibnamefont
  {Loudon}},\ }\href@noop {} {\bibfield  {journal} {\bibinfo  {journal}
  {Physical Review A}\ }\textbf {\bibinfo {volume} {68}},\ \bibinfo {pages}
  {013806} (\bibinfo {year} {2003})}\BibitemShut {NoStop}%
\bibitem [{\citenamefont {Schmiegelow}\ \emph {et~al.}(2016)\citenamefont
  {Schmiegelow}, \citenamefont {Schulz}, \citenamefont {Kaufmann},
  \citenamefont {Ruster}, \citenamefont {Poschinger},\ and\ \citenamefont
  {Schmidt-Kaler}}]{schmiegelow2016transfer}%
  \BibitemOpen
  \bibfield  {author} {\bibinfo {author} {\bibfnamefont {C.~T.}\ \bibnamefont
  {Schmiegelow}}, \bibinfo {author} {\bibfnamefont {J.}~\bibnamefont {Schulz}},
  \bibinfo {author} {\bibfnamefont {H.}~\bibnamefont {Kaufmann}}, \bibinfo
  {author} {\bibfnamefont {T.}~\bibnamefont {Ruster}}, \bibinfo {author}
  {\bibfnamefont {U.~G.}\ \bibnamefont {Poschinger}}, \ and\ \bibinfo {author}
  {\bibfnamefont {F.}~\bibnamefont {Schmidt-Kaler}},\ }\href@noop {} {\bibfield
   {journal} {\bibinfo  {journal} {Nature communications}\ }\textbf {\bibinfo
  {volume} {7}},\ \bibinfo {pages} {12998} (\bibinfo {year}
  {2016})}\BibitemShut {NoStop}%
\bibitem [{\citenamefont {Quinteiro}\ \emph {et~al.}(2017)\citenamefont
  {Quinteiro}, \citenamefont {Reiter},\ and\ \citenamefont
  {Kuhn}}]{quinteiro2017formulation}%
  \BibitemOpen
  \bibfield  {author} {\bibinfo {author} {\bibfnamefont {G.}~\bibnamefont
  {Quinteiro}}, \bibinfo {author} {\bibfnamefont {D.}~\bibnamefont {Reiter}}, \
  and\ \bibinfo {author} {\bibfnamefont {T.}~\bibnamefont {Kuhn}},\ }\href@noop
  {} {\bibfield  {journal} {\bibinfo  {journal} {Physical Review A}\ }\textbf
  {\bibinfo {volume} {95}},\ \bibinfo {pages} {012106} (\bibinfo {year}
  {2017})}\BibitemShut {NoStop}%
\bibitem [{\citenamefont {Scully}\ and\ \citenamefont
  {Zubairy}(1997)}]{scully1997qua}%
  \BibitemOpen
  \bibfield  {author} {\bibinfo {author} {\bibfnamefont {M.~O.}\ \bibnamefont
  {Scully}}\ and\ \bibinfo {author} {\bibfnamefont {M.~S.}\ \bibnamefont
  {Zubairy}},\ }\href@noop {} {\emph {\bibinfo {title} {Quantum Optics}}}\
  (\bibinfo  {publisher} {Cambridge University Press, Cambridge},\ \bibinfo
  {year} {1997})\BibitemShut {NoStop}%
\bibitem [{\citenamefont {Quinteiro}\ \emph {et~al.}(2010)\citenamefont
  {Quinteiro}, \citenamefont {Lucero},\ and\ \citenamefont
  {Tamborenea}}]{quinteiro2010ele}%
  \BibitemOpen
  \bibfield  {author} {\bibinfo {author} {\bibfnamefont {G.~F.}\ \bibnamefont
  {Quinteiro}}, \bibinfo {author} {\bibfnamefont {A.~O.}\ \bibnamefont
  {Lucero}}, \ and\ \bibinfo {author} {\bibfnamefont {P.~I.}\ \bibnamefont
  {Tamborenea}},\ }\href@noop {} {\bibfield  {journal} {\bibinfo  {journal}
  {J.\ Phys.\ Cond.\ Matter}\ }\textbf {\bibinfo {volume} {22}},\ \bibinfo
  {pages} {505802} (\bibinfo {year} {2010})}\BibitemShut {NoStop}%
\bibitem [{\citenamefont {Sauter}\ \emph {et~al.}(1986)\citenamefont {Sauter},
  \citenamefont {Neuhauser}, \citenamefont {Blatt},\ and\ \citenamefont
  {Toschek}}]{sauter1986observation}%
  \BibitemOpen
  \bibfield  {author} {\bibinfo {author} {\bibfnamefont {T.}~\bibnamefont
  {Sauter}}, \bibinfo {author} {\bibfnamefont {W.}~\bibnamefont {Neuhauser}},
  \bibinfo {author} {\bibfnamefont {R.}~\bibnamefont {Blatt}}, \ and\ \bibinfo
  {author} {\bibfnamefont {P.~E.}\ \bibnamefont {Toschek}},\ }\href {\doibase
  10.1103/PhysRevLett.57.1696} {\bibfield  {journal} {\bibinfo  {journal}
  {Phys. Rev. Lett.}\ }\textbf {\bibinfo {volume} {57}},\ \bibinfo {pages}
  {1696} (\bibinfo {year} {1986})}\BibitemShut {NoStop}%
\bibitem [{\citenamefont {Schmiegelow}\ and\ \citenamefont
  {Schmidt-Kaler}(2012)}]{schmiegelow2012light}%
  \BibitemOpen
  \bibfield  {author} {\bibinfo {author} {\bibfnamefont {C.}~\bibnamefont
  {Schmiegelow}}\ and\ \bibinfo {author} {\bibfnamefont {F.}~\bibnamefont
  {Schmidt-Kaler}},\ }\href@noop {} {\bibfield  {journal} {\bibinfo  {journal}
  {European Physical Journal D}\ }\textbf {\bibinfo {volume} {66}},\ \bibinfo
  {pages} {1} (\bibinfo {year} {2012})}\BibitemShut {NoStop}%
\end{thebibliography}
\end{document}